\documentclass[letterpaper]{article} %
\usepackage[]{aaai25}  %
\usepackage{times}  %
\usepackage{helvet}  %
\usepackage{courier}  %
\usepackage[hyphens]{url}  %
\usepackage{graphicx} %
\usepackage{natbib}  %
\usepackage{caption} %
\usepackage{algorithm}
\usepackage{algorithmic}
\usepackage{xcolor}

\usepackage{newfloat}
\usepackage{listings}
\DeclareCaptionStyle{ruled}{labelfont=normalfont,labelsep=colon,strut=off} %
\floatstyle{ruled}
\newfloat{listing}{tb}{lst}{}
\floatname{listing}{Listing}
\title{AffectMachine-Pop: A controllable expert system for real-time pop music generation}
\author{}
\author{
    Kat R. Agres\textsuperscript{\rm 1}\equalcontrib,
    Adyasha Dash\textsuperscript{\rm 2}\equalcontrib ,
    Phoebe Chua\textsuperscript{\rm 3},
    and Stefan K. Ehrlich\textsuperscript{\rm 4}
}
\affiliations{
    \textsuperscript{\rm 1}Centre for Music and Health, Yong Siew Toh Conservatory of Music, National University of Singapore, Singapore, Singapore\\
    \textsuperscript{\rm 2}Yong Siew Toh Conservatory of Music, National University of Singapore, Singapore, Singapore\\
    \textsuperscript{\rm 3}Department of Information Systems and Analytics, National University of Singapore, Singapore, Singapore\\
    \textsuperscript{\rm 4}SETLabs Research GmbH, Munich, Germany\\

}

\begin{document}

\maketitle

\begin{abstract}
Music is a powerful medium for influencing listeners' emotional states, and this capacity has driven a surge of research interest in AI-based affective music generation in recent years. Many existing systems, however, are a black box which are not directly controllable, thus making these systems less flexible and adaptive to users.
We present \textit{AffectMachine-Pop}, an expert system capable of generating retro-pop music according to arousal and valence values, which can either be pre-determined or based on a listener's real-time emotion states.
To validate the efficacy of the system, we conducted a listening study demonstrating that AffectMachine-Pop is capable of generating affective music at target levels of arousal and valence. The system is tailored for use either as a tool for generating interactive affective music based on user input, or for incorporation into biofeedback or neurofeedback systems to assist users with emotion self-regulation.
\end{abstract}

\section{Introduction}
Artificial intelligence (AI)-based music generation has rapidly gained traction in recent years, driven by advances in machine learning, computational creativity, and the growing availability of music datasets. Such systems have broad applications, from enhancing artistic creativity and aiding music composition through co-creative systems, to providing interactive entertainment in gaming and VR scenarios, and even person-centered, adaptive therapeutic applications \cite{ji2020comprehensive,agres2021music,dash2024ai, civit2022systematic, wang2024review}.

Advancements in generative AI, specifically large language models (LLM) and natural language processing (NLP) technology, have led to groundbreaking advancements in the field. Generative models such as AudioLM \cite{borsos2023audiolm}, MusicLM \cite{agostinelli2023musiclm}, and MusicGen \cite{copet2024simple} can be prompted with either text descriptions or melodies to produce high-quality audio samples that mimic diverse genres, styles and sentiments. Aside from text-based prompting, previous approaches have also achieved music generation by sampling latent spaces encoding distinct musical attributes, styles, and sentiments \cite{tan2020music, gupta2023towards}.

Within the field of music generation, there has been significant interest in AI-based \textit{affective} music generation (AI-AMG) systems, which aim to create music with specific emotional qualities \cite{grekow2021monophonic, sulun2022symbolic, cui2024moodloopgp, dash2024ai}. Through their ability to induce, mediate and enhance listener’s emotional states, AI-AMG systems widen the scope of algorithmic music generation by enabling tailored emotional engagement in applications such as mental health interventions \cite{elliott2011relaxing, stewart2019music,agres2021music}, assistive technology \cite{hagerer2015augmenting}, and physical rehabilitation \cite{ashoori2015effects}.

While generative AI has brought about important advancements in the development of AI-AMGs, specifically regarding the high quality and naturalness of the generated music, most approaches relying on generative AI (especially neural network-based approaches) suffer from several drawbacks, namely: 1) the generation of musical samples from generative AI models is typically performed at the song level and usually requires several seconds up to minutes; 2) near real-time inference for instantaneous music generation is rare; and 3) the music generation process is non-deterministic, rendering fine-grained control of the affective qualities of the music challenging, and the expected emotional impact on the listener non-trivial. Furthermore, the majority of generative AI models rely on pre-existing music for training, which has sparked controversies surrounding copyrighted training data, making approaches \textit{not} relying on pre-existing music increasingly attractive.

These factors render state-of-the-art neural networks-based generative AI music systems sub-optimal for real-time music generation, which is essential for interactive entertainment in gaming and VR scenarios, as well as human-in-the-loop adaptive therapeutic systems. The challenge of \textit{narrative adaptability} \cite{dash2024ai} -- the ability of AI-AMG systems to instantaneously generate music seamlessly adapting to a given emotional narrative or sequence, while still maintaining musical coherence -- has not been explored in great depth. However, real-time adaptation of musical output, which can in turn influence the listener's emotions, is a key area given music's significant impact on human emotion \citep{juslin2013music}. 
Few systems have addressed this challenge. For example, \citet{engels2015automatic} employed hierarchical Markov models to generate original music for video games, capable of mimicing the variations in musical structure often found in longer human-composed pieces. Similarly, \citet{ehrlich2019closed} developed a rule-based probabilistic music generation algorithm as part of a closed-loop brain-computer interface system. The algorithm generates a stream of MIDI events, with the type and occurrence of events modulated by emotional arousal and valence parameters. Recent work by \citet{agres2023affectmachine} introduced an AI-AMG system called \textit{AffectMachine-Classical}, capable of creating non-repetitive, continuous affective music in a classical style. This system was validated in a study with human participants, confirming the emotional qualities of the generated musical and supporting its utility for the aforementioned use cases.

Building upon AffectMachine Classical, we present here \textit{AffectMachine-Pop}, a novel AI-AMG system designed to generate continuous affective music in a pop style.
Compared with classical music, pop music is characterized by simpler, more repetitive forms that resonate more broadly with the general public \cite{chan2009visualizing}. However, generating multi-track pop music presents the technical challenge of maintaining melodic, harmonic and rhythmic coherence across several musical layers \cite{ren2020popmag, zhu2018xiaoice}. Therefore, unlike most modern pop music AI-AMGs, which predominantly rely on data-driven methods and high-performance AI architectures, e.g. \cite{ren2020popmag, huang2020pop}, our approach takes a different path. Existing pop systems often depend heavily on pre-existing music datasets and suffer from limited transparency, functioning as black-box models. In contrast, our method features a compositional framework guided by established musical rules, deliberate instrumentation, and precise timbral arrangements. These elements are manually refined by expert musicians and composers to achieve a balance of technical accuracy and artistic expressiveness. In addition, our approach features near real-time adaptability, enabling the system to seamlessly adjust its affective content based on the user's inputs or physiological states, by traversing the valence-arousal space \citep{russell1980circumplex} while preserving musical and stylistic coherence. The system generates musical samples with controllable emotional qualities, i.e., music at specific levels of arousal and valence, as demonstrated by empirical evidence from the listening study presented here.

Below we describe \textit{AffectMachine-Pop}, our expert system for affective automatic music generation in a popular music style. We also highlight the ways in which the system compares and differs from AffectMachine-Classical. Finally, we present the methodology and results of a listening study conducted to validate the system's ability to produce music at different target levels of arousal and valence.

\section{AffectMachine-Pop system description} \label{AMGS}
Here we describe the parameters and design of our system, \textit{AffectMachine-Pop}, for computationally-generating affective pop music. Unlike the majority of AI-AMG systems, which have focused on generating music in a classical genre (e.g., \citet{agres2023affectmachine, ehrlich2019closed, wallis2011rule}), the present system composes music in a retro-pop genre. This makes it particularly well-suited for middle-aged and older adults, and anyone who enjoys popular music from the 60s and 70s. 

\begin{figure*}
    \centering
    \includegraphics[width=0.9\linewidth]{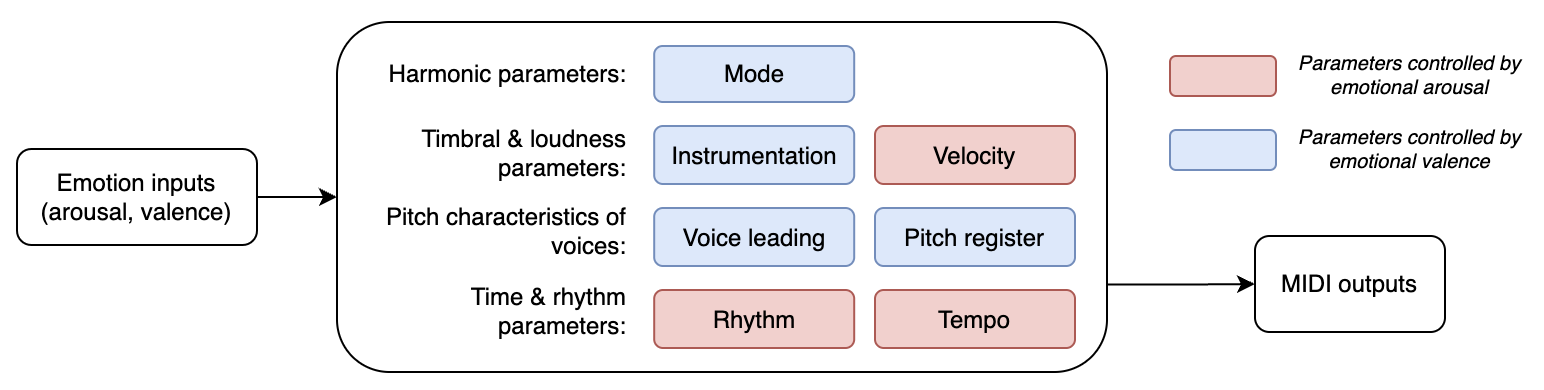}
    \caption{Architecture of AffectMachine-Pop.}
    \label{fig:system-figure}
\end{figure*}

Figure \ref{fig:system-figure} depicts the basic building blocks of our system. There are four major components, namely, \textit{Harmonic parameters}, \textit{Timbral \& loudness parameters}, \textit{Pitch characteristics of voices}, and \textit{Time \& rhythm parameters}. Each of these blocks contains different musical rules for tailoring features of the music (i.e., tempo, chord, etc) to convey specific levels of emotion. The emotion parameters (valence and arousal values) are given as input to the system, and then the system generates emotional music (output) through a complex interplay between these blocks. In the following section, we describe each of these blocks in greater detail.

 We focus on highlighting the \textit{differences} between AffectMachine-Pop and AffectMachine-Classical, which has already been described in-depth by \citet{agres2023affectmachine}. Where the designs overlap, we offer only a brief summary.

\subsection{Emotion inputs}
Russell’s (1980) two-dimensional model of affective space, which represents emotions along the dimensions of arousal (intensity or energy level) and valence (degree of pleasantness) \citep{russell1980circumplex}, has inspired many AI-AMG systems \cite{dash2024ai, huang2024emotion, agres2023affectmachine, ehrlich2019closed}, including \textit{AffectMachine-Pop}. This model has been widely adopted (and adapted) in emotion research, as it allows for the representation of a broad range of emotions, from high-arousal, positive states (e.g., joy) to low-arousal, negative states (e.g., sadness). 
By mapping musical features to this affective space, AffectMachine-Pop can generate music that meets target arousal and valence levels, which are either predetermined, or based on real-time inputs, such as physiological data reflecting a listener’s emotion states (such as HR or EEG data).

\subsection{Harmonic parameters}
\subsubsection{Mode}
\begin{figure}[h!]
    \includegraphics[width=\linewidth]{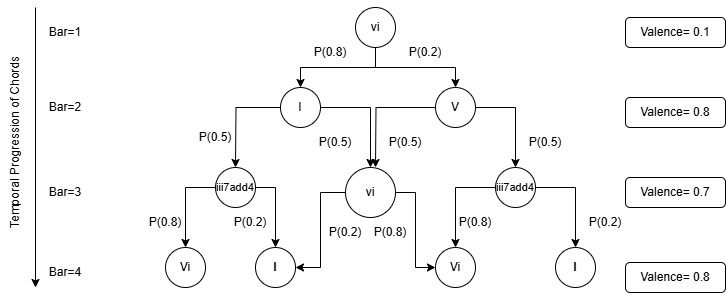}
    \caption{Sample Chord Progression as a probabilistic graph.}
    \label{fig: Sample Chord Progression as a probabilistic graph}
\end{figure}

AffectMachine-Pop uses a bespoke probabilistic chord progression matrix that takes inspiration from retro-pop music, e.g., English and Mandarin pop songs from the 1960s-1970s, such as The Beatles’ \textit{Hey Jude} and Theresa Teng’s \textit{Thousands of Words}. Specifically, the chord progression matrix can be modeled as a directed probabilistic graph, $G(V, E)$ with vertices $V$ and edges $E$. Each vertex represents a chord, and edges represent transition probabilities between chords. The generated chord sequence is guided by a pre-defined sequence or array of valence values; each valence value corresponds to one bar in the progression, such that the array's length matches the number of bars in the generated music.
The chords and probabilities in our expert system were tuned by hand based on previous empirical findings, as well as the musical insights from our interdisciplinary team of musicians and composers. Figure \ref{fig: Sample Chord Progression as a probabilistic graph} shows a representation of our chord progression algorithm as a directed probability graph for a 4-bar musical excerpt with a valence parameter of [0.1, 0.8, 0.7, 0.8]. 

The musical structure governing the length of the generated chord progressions differs from that of AffectMachine-Classical. In AffectMachine-Pop, the music follows a 32-bar musical structure comprised of two 8-bar sections (labelled section “A” and “B”, as per convention) that repeat in an AABB pattern. (This AABB pattern is distinct from the AffectMachine-Classical system.) Each bar in the structure is associated with a fixed chord function, while the specific chords for the bar and their associated transition probabilities are determined by the valence level.

\subsection{Timbral and loudness parameters}

\subsubsection{Instrumentation}
AffectMachine-Pop uses five virtual instruments, played over different audio tracks, to generate affective polyphonic music that resembles ``retro'' pop music from the 1960s-1970s. The five instruments are a percussion kit, bass guitar, electric guitar, violin section and French horn. Each instrument is played using a single audio track in the DAW (Ableton Live) except for the electric guitar, which is played back over two audio tracks. The violin section and French horn receive identical MIDI information and play the main melodic line. The doubling of the violin section by French horn helps to smooth and balance the timbre and enhance the prominence of the melodic line. The electric guitar provides harmonic accompaniment by strumming chords, which are played back on one audio track. At lower levels of arousal (aro $<$0.7) where the music is slower and less dense with minimal percussion, a plucked electric guitar is also used to add texture, to convey a desired level of valence when note activations were very sparse. The strummed and plucked electric guitars are recorded on separate audio tracks to facilitate audio mixing.

\subsubsection{Velocity}
Volume (or dynamics) is frequently used to convey emotional expressiveness in music \cite{gabrielsson2003emotional}. Increased sound levels are associated with higher-arousal emotions such as ``happy" or ``angry'', while lower sound levels tend to be associated with lower-arousal emotions such as ``sad" \cite{bresin2011emotion}. To enable real-time adaptation of volume, we mapped arousal values to attack velocity in MIDI following Eq. \ref{eq:loudness}. Attack velocity indicates the force with which a note is struck; aside from controlling volume, higher velocity values also produce brighter and harder timbres which contribute to the perception of increased arousal \cite{agres2023affectmachine}.

\begin{equation}\label{eq:loudness}
    velocity = 60 + (aro*15)
\end{equation}

\subsection{Pitch characteristics of voices}
\subsubsection{Voice leading}
Voice leading -- the linear progression of melodic voices and their interaction to create harmonies -- was added to the system for musical sophistication and coherence.
Separate voice leading logic was developed for each instrument and track in AffectMachine-Pop. 
The bass guitar plays the roots and fifths of the current chord, a common note pattern used by bass players in pop music. The root is selected with higher probability ($p=0.9$) on the first beat of each bar to help establish the chord. Next, the plucked electric guitar is used primarily to add texture at lower levels of arousal; hence, it plays a randomly selected sequence of chord tones from the current chord (with all chord tones being equiprobable). This allows the current chord quality to be expressed clearly even at lower levels of arousal with little additional dissonance.

The strummed electric guitar provides harmonic progression, and its voice leading logic follows the heuristic outlined in \citet{wallis2011rule}, which is that musicians tend to voice new chords in a manner that is as similar as possible to the previous chord, in terms of interval and placement on the instrument. We implement this by selecting the voicing for the subsequent chord that is least dissimilar to the current chord (details of the dissimilarity calculation can be found in \citet{agres2023affectmachine}).

The main melodic line is played by the violin section and French horn. For these instruments, we employ a mix of composed melodic motifs and probabilistic voice leading logic. Specifically, the instruments play a melodic pattern probabilistically determined by the voice leading logic for the first four bars of each 8-bar section. Like AffectMachine-Classical, this voice leading logic is encoded in the form of transition matrices. However, in AffectMachine-Pop, we divide the valence range into low ($V \leq 0.5$) and high valence ($V>0.5$) regions and compose one matrix for each region to generate appropriate melodies for each level of valence. In the second four bars, the instruments play a probabilistically-selected composed rhythmic motif. The set of composed rhythmic motifs were developed with reference to representative pieces from the Western and mandarin retro-pop music canon during the 1960s-1980s, and include typical motifs such as pentatonic patterns and arpeggiation.

\subsubsection{Pitch register}
Generally, higher pitches tend to be associated with positive emotions such as excitement and serenity \cite{collier1998judgments}, while lower pitches correlate with negatively-valenced emotions such as sadness. 
We use chord inversions to gradually shift the pitch register of the strummed and plucked electric guitar. Regardless of the current level of valence, there is always a constant probability ($p=0.6$) that the chord voicing does not change, to prevent the pitch register of the music from increasing or decreasing too quickly. As valence increases, the probability that the inversion of the current chord voicing increases by one (e.g., from root to first inversion) also increases according to the formula $p(inv+ | inv \neq 0) = val$, while the probability that the inversion of the current chord voicing decreases by one (e.g., from first to root inversion) decreases according to the formula $p(inv- | inv \neq 0) = 1-val$. 

Register is not relevant for the percussion instrument, and for the remaining instruments (French horn and violin section), the register is fairly consistent as they return to the composed melodic motive in the second four bars of each 8-bar section, and there are typically no obvious register changes due to voice leading logic in the remaining four bars.

\subsection{Time and rhythm parameters}
\subsubsection{Rhythm}
We composed a set of three equiprobable 8-bar rhythmic patterns for the bass guitar, and randomly select a new rhythmic pattern at the beginning of each 8-bar section. For the strummed electric guitar, we divided the arousal range into three regions:  low ($aro<0.40$), moderate ($0.4 \leq aro < 0.70$) and high ($aro \geq 0.70$), and composed several rhythmic patterns (and their associated probabilities) for each region. A new rhythmic pattern is selected for strummed guitar at the beginning of each bar. For the percussion kit, we composed several 8-bar rhythmic patterns that gradually reduce in density as arousal decreases, with greater use of quieter instruments and techniques such as rim clicks. A new rhythmic pattern for percussion is selected at the beginning of each 8-bar section. A single pattern was composed for low ($aro \leq 0.30$) and moderate ($0.3 < aro \leq 0.70$) levels of arousal, while a set of three rhythmic patterns were composed for high ($aro > 0.70$) levels of arousal.

For the main melodic instruments (string section and French horn), we employ a mix of composed rhythmic motives and rhythmic roughness. Specifically, during the first four bars of each 8-bar section, the instruments play a rhythmic pattern probabilistically determined by the roughness parameter, which is a measure of how irregular the rhythm of a piece of music is. Music with smooth, regular rhythms are typically perceived as higher in valence and lower arousal. In AffectMachine-Pop, we use note density as a proxy for rhythmic roughness \cite{wallis2011rule} in the following manner: as arousal increases, roughness decreases in a linear fashion and note density increases. For example, if roughness is set to 0, each bar would be populated with eight notes of equal length. Because this tends to sound overly dense (since the tempo is also faster at higher levels of arousal), we set a lower bound for the roughness parameter. 

In terms of the last four bars of each 8-bar section, the instruments play a selected composed rhythmic motive. We divided the arousal range into three regions: low ($aro < 0.30$), moderate ($0.3 \leq aro < 0.60$) and high ($aro \geq 0.60$), and composed a set of three equiprobable rhythmic patterns for each region. Finally, rhythmic patterns for the plucked electric guitar are determined probabilistically by the roughness parameter.

\subsubsection{Tempo}
Tempo refers to the rate or speed of the music, often measured in terms of beats-per-minute (bpm). In our system, tempo is controlled by the arousal value. Specifically, tempo is governed by a logarithmic relationship with arousal, and has a range of $tempo \in [36,130]$. We found the logarithmic relationship and limited range to be useful, as changes in tempo were more perceptually apparent when the tempo was lower, and using overly fast tempos sounded unpleasant and unnatural.

\section{AffectMachine-Pop Listening study} \label{Study}
To evaluate the effectiveness of AffectMachine-Pop in generating music at specific levels of arousal and valence, we conducted a music listening study. This study assesses the system's ability to convey the intended emotional content of the generated music to listeners (i.e., perceived emotion). Future work will examine AffectMachine's capacity to reliably induce emotions in listeners.

\subsubsection{Musical Stimuli Generation}
AffectMachine-Pop is designed to compose music that conveys emotions corresponding  to any point on the valence-arousal (V-A) plane. To capture a range of different emotions, we generated musical excerpts from 13 different points on the V-A plane. These points were selected to span a variety of different emotions, including the corners, the middle of each quadrant, and the neutral point at the centre of the space. For more details on the selection of points in A-V space, please refer to \citet{agres2023affectmachine}. 
To minimize bias towards any single musical excerpt and improve generalizability, three musical stimuli were generated for each of these 13 points, resulting in a total of  39 musical excerpts. The average duration of the excerpts was 32.6 seconds. The musical stimuli were composed based on a 4-bar, 8-bar, or 16-bar progression. For stimuli corresponding to low arousal levels, the duration was limited to 4 bars (ranging from 27-34 sec per stimulus).  Stimuli with higher levels of arousal contained 8 or 16 bars, as 4 bars were too brief for these stimuli with a faster tempo.

\subsubsection{Experimental Procedure}

We conducted the listening study with 24 participants (average age = 23.0 yrs, SD = 2.6 yrs; male = 9, female = 15). All participants were given verbal and written instructions about the study prior to providing their written consent. The study was approved by the Institutional Review Board (IRB) of the National University of Singapore (NUS).

The experiment took place in a quiet room free from audio-visual distractions. The experimenter explained the procedure to each participant, who then provided written informed consent to participate in the study. Participants completed the study individually. Before starting the listening task, participants provided demographic information such as their age, gender, and ethnicity.

At the beginning of the study, participants listened to two practice trials to become familiar with the procedure. Subsequently, the 39 musical excerpts were presented in randomized order. After listening to each stimulus, participants were asked to indicate their perceived emotion, i.e., the emotion they perceived in the musical excerpt. Responses were collected using the Self-Assessment Manikin (SAM), a visual 9-point scale ranging from ``Very unpleasant'' (1) to ``Extremely pleasant'' (9) for valence, and from ``Calm'' (1) to ``Excited'' (9) for arousal. Participants were allowed to listen to each stimulus only once, but could take as much time as needed to provide their ratings. The experiment lasted approximately 40 minutes, and participants were compensated with \$ 6 SGD for their time.

The valence and arousal ratings collected across participants were then analyzed to evaluate the efficacy of our system in conveying specific emotions through its generated music. The results of the analyses are presented below.

\section{Results and Discussion} \label{ResDis}

To evaluate the efficacy of AffectMachine-Pop in generating music that expresses a desired emotion, we analyzed the user ratings collected during the listening study. The primary goal was to determine whether the music generated by AffectMachine-pop successfully conveys the intended levels of valence and arousal to listeners. Specifically, we performed a comparative analysis between the average user ratings for the musical stimuli and the valence or arousal parameter settings used during the music generation process. 
First, we normalized the perceptual ratings for each listener using Eq. 2. Here, the $Max_{Valence}$ refers to the maximum possible valence rating (i.e., 9) and $Min_{Valence}$ refers to the minimum possible valence rating (i.e., 1). A similar normalization procedure was applied for arousal. The resulting normalized valence and arousal ratings, which range from 0 to 1, were used for further analysis. In the remainder of the article, the normalized valence and normalized arousal ratings will be referred to as valence and arousal ratings, respectively.

\begin{equation}\label{eqn:Normalized_Valence}
Normalized_{Valence} = \frac{Rated_{Valence}- Min_{Valence}}{(Max_{Valence} – Min_{Valence})}
\end{equation}

\subsubsection{Comparative Analysis of User Ratings}

\begin{figure}[h!]
    \includegraphics[width =0.475\textwidth]{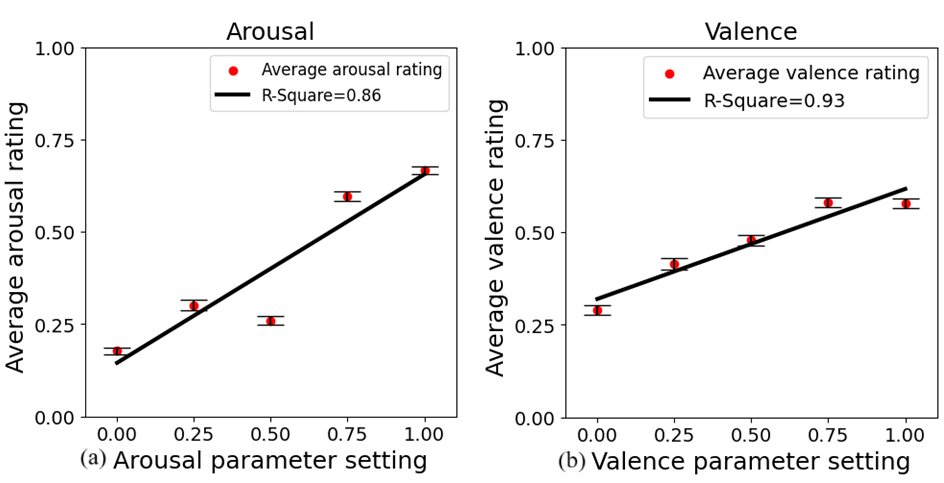}
    \caption{Mean arousal (a) and valence (b) ratings (error bars represent standard error).}
    \label{fig: Regression Plot}
\end{figure}

To access whether our system accurately expresses the desired emotions through its generated music, we performed a comparative analysis between the mean emotion ratings and the valence and arousal settings used to generate the music. For example, the mean arousal ratings for stimuli generated with settings {valence, arousal} = [\{0,1\}, \{0.5,1\}, \{1,1\}] were used to evaluate the system’s performance when arousal was set to its maximum value.  Figure \ref{fig: Regression Plot} presents the average ratings along with their respective standard errors. 

As shown in the figure, the valence ratings exhibit a positive linear trend but tend to plataeu when $V >= 0.75$. According to literature, the middle range of psychometric rating scales often receives a higher density of responses, whereas the extremes of the scale typically receive fewer responses \cite{leung2011comparison}. This phenomenon could explain why raters found extreme values of valence, such as valence values $> 0.75$ and $< 0.25$, to be less distinguishable. 

In contrast, arousal values showed better correspondence with the average user ratings across most of the scale, except around an arousal value of 0.5. This may be due to the fact that the excerpts were generated using three random instantiations (e.g., these particular stimuli may have deviated from the system's typical output at $A = 0.5$). It is also possible that the tempo around $A = 0.5$ is too slow. Future work will explore a faster tempo for moderate levels of arousal, and will test a larger number of excerpts. 

To quantify the positive correlation between average user ratings and parameter settings, we performed linear regression-based curve fitting. The coefficient of determination was $R^2 = 0.93$ $(p < 0.01)$ for valence and $R^2 = 0.86$ $(p < 0.05)$ for arousal, indicating good correspondence between user ratings and target values and thereby confirming the effectiveness of our system to convey the target emotion.

\begin{figure}[h!]
    \includegraphics[width=0.475\textwidth]{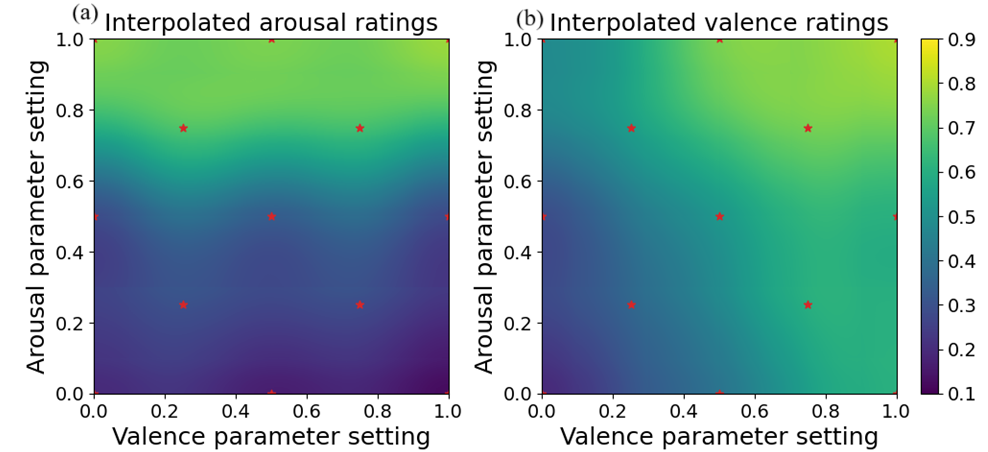}
    \caption{Mean (interpolated) arousal (a) and valence (b) ratings as a function of the valence and arousal parameters. Vertical color bars represent the colors
 corresponding to different values of normalized average ratings over the range 0.1 to 0.9.}
    \label{fig: Interpolation Plot}
\end{figure}

We also investigated whether the perception of arousal is influenced by variations in valence and vice versa. To this end, we examined the dependence of perceived emotion user ratings on both the valence and arousal parameter settings simultaneously. Figure \ref{fig: Interpolation Plot} illustrates this relationship, showing the interpolated average arousal ratings (Fig 4a) and valence ratings (Fig 4b). The stars correspond to the 13 points in the valence-arousal plane used to generate the musical stimuli. 

Generally, there is a strong correspondence between parameter settings and emotion ratings.
As shown in the figure (Fig 4b), perceived valence is lower than the intended valence parameter setting (for values of $V > 0.8$) when arousal is set below $< 0.6$. This implies that excerpts intended to convey high valence are perceived as having only moderate valence when arousal value is low. Conversely, stimuli generated at low valence settings (values of $V < 0.2$) are perceived to express higher valence when arousal is set high (values of $A > 0.7$). This suggests that music intended to express low valence is perceived as having higher valence when the arousal value is higher. These observations can be partially attributed to the effect of tempo in generated music, where slower tempos are associated with lower arousal values and vice versa. 

In contrast, we observed a more consistent correspondence between the arousal parameter settings and arousal ratings (except for $A = 0.5$), and found that the arousal ratings show no dependence on valence settings. This observation aligns with findings from the literature \cite{wallis2011rule}, which reported an asymmetrical ``crossover'' effect between arousal and valence. Specifically, \citet{wallis2011rule} observed that while perceived valence correlates with intended arousal, perceived arousal does not significantly correlate with intended valence. 
The observed crossover effect is most likely due to the non-orthogonal relationship between the valence and arousal dimensions in relation to one or more musical features. Alternatively, it could also arise from uncontrollable factors, such as cultural influences. Nevertheless, this crossover effect does not greatly impact the intended functioning of our systems, as the results within each dimension align well with expectations.

These findings confirm the potential of AffectMachine-Pop to reliably generate emotion-infused music that expresses varying levels of arousal and valence in a controlled manner. However, we note that the results may be influenced by the limited sample size, which could affect generalizability. Further in-depth investigations will be useful to expand on the above observations.

\section{Conclusion} \label{Con}
We have presented \textit{AffectMachine-Pop}, an expert system designed for generating affective music in a pop style. This system is capable of real-time music generation, offering a novel and innovative tool for creating music tailored to convey specific emotion states.

In a music listening study, we investigated whether the system is able to generate music that effectively expresses desired emotions (specified via specific levels of arousal and valence). The results showed a strong correspondence between the given valence and arousal parameter settings and listeners' averaged valence and arousal values, with $R^2 = 0.93$ for valence and $R^2 = 0.86$ for arousal. These results confirm the system's effectiveness.

Future work will evaluate the system's ability to \textit{induce} emotions in listeners, extending beyond the current focus on perceived emotion. Additionally, we aim to address current limitations, including the need for a larger and more diverse group of participants. 
The long-term goals of this line of research are to establish the system's potential to reliably generate affective music, and explore its applications in emotion self-regulation. One promising future direction is the integration of AffectMachine into a Brain-Computer-Interface (BCI), which aims to assist users in achieving desired emotional or mood states through adaptive, real-time music generation guided by their neural activity.

\section{Acknowledgments}
This work was supported by the RIE2020 Advanced Manufacturing and Engineering (AME) Programmatic Fund (No. A20G8b0102), Singapore. We wish to thank the composer Cliff Tan for his musical contributions to the development of AffectMachine-Pop, and our former research assistant, Ting Yuan Tay, for her help with data collection for  the listener study.

\bibliography{refs}

\begin{thebibliography}{32}
\providecommand{\natexlab}[1]{#1}

\bibitem[{Agostinelli et~al.(2023)Agostinelli, Denk, Borsos, Engel, Verzetti, Caillon, Huang, Jansen, Roberts, Tagliasacchi et~al.}]{agostinelli2023musiclm}
Agostinelli, A.; Denk, T.~I.; Borsos, Z.; Engel, J.; Verzetti, M.; Caillon, A.; Huang, Q.; Jansen, A.; Roberts, A.; Tagliasacchi, M.; et~al. 2023.
\newblock Musiclm: Generating music from text.
\newblock \emph{arXiv preprint arXiv:2301.11325}.

\bibitem[{Agres, Dash, and Chua(2023)}]{agres2023affectmachine}
Agres, K.~R.; Dash, A.; and Chua, P. 2023.
\newblock AffectMachine-Classical: a novel system for generating affective classical music.
\newblock \emph{Frontiers in Psychology}, 14: 1158172.

\bibitem[{Agres et~al.(2021)Agres, Schaefer, Volk, van Hooren, Holzapfel, Dalla~Bella, M{\"u}ller, De~Witte, Herremans, Ramirez~Melendez et~al.}]{agres2021music}
Agres, K.~R.; Schaefer, R.~S.; Volk, A.; van Hooren, S.; Holzapfel, A.; Dalla~Bella, S.; M{\"u}ller, M.; De~Witte, M.; Herremans, D.; Ramirez~Melendez, R.; et~al. 2021.
\newblock Music, computing, and health: a roadmap for the current and future roles of music technology for health care and well-being.
\newblock \emph{Music \& Science}, 4: 2059204321997709.

\bibitem[{Ashoori, Eagleman, and Jankovic(2015)}]{ashoori2015effects}
Ashoori, A.; Eagleman, D.~M.; and Jankovic, J. 2015.
\newblock Effects of auditory rhythm and music on gait disturbances in Parkinson’s disease.
\newblock \emph{Frontiers in neurology}, 6: 234.

\bibitem[{Borsos et~al.(2023)Borsos, Marinier, Vincent, Kharitonov, Pietquin, Sharifi, Roblek, Teboul, Grangier, Tagliasacchi et~al.}]{borsos2023audiolm}
Borsos, Z.; Marinier, R.; Vincent, D.; Kharitonov, E.; Pietquin, O.; Sharifi, M.; Roblek, D.; Teboul, O.; Grangier, D.; Tagliasacchi, M.; et~al. 2023.
\newblock Audiolm: a language modeling approach to audio generation.
\newblock \emph{IEEE/ACM transactions on audio, speech, and language processing}, 31: 2523--2533.

\bibitem[{Bresin and Friberg(2011)}]{bresin2011emotion}
Bresin, R.; and Friberg, A. 2011.
\newblock Emotion rendering in music: range and characteristic values of seven musical variables.
\newblock \emph{cortex}, 47(9): 1068--1081.

\bibitem[{Chan, Qu, and Mak(2009)}]{chan2009visualizing}
Chan, W.-Y.; Qu, H.; and Mak, W.-H. 2009.
\newblock Visualizing the semantic structure in classical music works.
\newblock \emph{IEEE transactions on visualization and computer graphics}, 16(1): 161--173.

\bibitem[{Civit et~al.(2022)Civit, Civit-Masot, Cuadrado, and Escalona}]{civit2022systematic}
Civit, M.; Civit-Masot, J.; Cuadrado, F.; and Escalona, M.~J. 2022.
\newblock A systematic review of artificial intelligence-based music generation: Scope, applications, and future trends.
\newblock \emph{Expert Systems with Applications}, 209: 118190.

\bibitem[{Collier and Hubbard(1998)}]{collier1998judgments}
Collier, W.~G.; and Hubbard, T.~L. 1998.
\newblock Judgments of happiness, brightness, speed and tempo change of auditory stimuli varying in pitch and tempo.
\newblock \emph{Psychomusicology: A Journal of Research in Music Cognition}, 17(1-2): 36.

\bibitem[{Copet et~al.(2024)Copet, Kreuk, Gat, Remez, Kant, Synnaeve, Adi, and D{\'e}fossez}]{copet2024simple}
Copet, J.; Kreuk, F.; Gat, I.; Remez, T.; Kant, D.; Synnaeve, G.; Adi, Y.; and D{\'e}fossez, A. 2024.
\newblock Simple and controllable music generation.
\newblock \emph{Advances in Neural Information Processing Systems}, 36.

\bibitem[{Cui, Sarmento, and Barthet(2024)}]{cui2024moodloopgp}
Cui, W.; Sarmento, P.; and Barthet, M. 2024.
\newblock MoodLoopGP: Generating Emotion-Conditioned Loop Tablature Music with Multi-granular Features.
\newblock In \emph{International Conference on Computational Intelligence in Music, Sound, Art and Design (Part of EvoStar)}, 97--113. Springer.

\bibitem[{Dash and Agres(2024)}]{dash2024ai}
Dash, A.; and Agres, K. 2024.
\newblock AI-Based Affective Music Generation Systems: A Review of Methods and Challenges.
\newblock \emph{ACM Computing Surveys}, 56(11): 1--34.

\bibitem[{Ehrlich et~al.(2019)Ehrlich, Agres, Guan, and Cheng}]{ehrlich2019closed}
Ehrlich, S.~K.; Agres, K.~R.; Guan, C.; and Cheng, G. 2019.
\newblock A closed-loop, music-based brain-computer interface for emotion mediation.
\newblock \emph{PloS one}, 14(3): e0213516.

\bibitem[{Elliott, Polman, and McGregor(2011)}]{elliott2011relaxing}
Elliott, D.; Polman, R.; and McGregor, R. 2011.
\newblock Relaxing music for anxiety control.
\newblock \emph{Journal of music therapy}, 48(3): 264--288.

\bibitem[{Engels, Tong, and Chan(2015)}]{engels2015automatic}
Engels, S.; Tong, T.; and Chan, F. 2015.
\newblock Automatic real-time music generation for games.
\newblock In \emph{Proceedings of the AAAI Conference on Artificial Intelligence and Interactive Digital Entertainment}, volume~11, 220--222.

\bibitem[{Gabrielsson and Juslin(2003)}]{gabrielsson2003emotional}
Gabrielsson, A.; and Juslin, P.~N. 2003.
\newblock \emph{Emotional expression in music.}
\newblock Oxford University Press.

\bibitem[{Grekow and Dimitrova-Grekow(2021)}]{grekow2021monophonic}
Grekow, J.; and Dimitrova-Grekow, T. 2021.
\newblock Monophonic music generation with a given emotion using conditional variational autoencoder.
\newblock \emph{IEEE Access}, 9: 129088--129101.

\bibitem[{Gupta et~al.(2023)Gupta, Kamath, Wei, Li, Nanayakkara, and Wyse}]{gupta2023towards}
Gupta, C.; Kamath, P.; Wei, Y.; Li, Z.; Nanayakkara, S.; and Wyse, L. 2023.
\newblock Towards controllable audio texture morphing.
\newblock In \emph{ICASSP 2023-2023 IEEE International Conference on Acoustics, Speech and Signal Processing (ICASSP)}, 1--5. IEEE.

\bibitem[{Hagerer et~al.(2015)Hagerer, Lux, Ehrlich, and Cheng}]{hagerer2015augmenting}
Hagerer, G.~J.; Lux, M.; Ehrlich, S.; and Cheng, G. 2015.
\newblock Augmenting affect from speech with generative music.
\newblock In \emph{Proceedings of the 33rd Annual ACM Conference Extended Abstracts on Human Factors in Computing Systems}, 977--982.

\bibitem[{Huang, Chen, and Yang(2024)}]{huang2024emotion}
Huang, J.; Chen, K.; and Yang, Y.-H. 2024.
\newblock Emotion-driven Piano Music Generation via Two-stage Disentanglement and Functional Representation.
\newblock \emph{arXiv preprint arXiv:2407.20955}.

\bibitem[{Huang and Yang(2020)}]{huang2020pop}
Huang, Y.-S.; and Yang, Y.-H. 2020.
\newblock Pop music transformer: Beat-based modeling and generation of expressive pop piano compositions.
\newblock In \emph{Proceedings of the 28th ACM international conference on multimedia}, 1180--1188.

\bibitem[{Ji, Luo, and Yang(2020)}]{ji2020comprehensive}
Ji, S.; Luo, J.; and Yang, X. 2020.
\newblock A comprehensive survey on deep music generation: Multi-level representations, algorithms, evaluations, and future directions.
\newblock \emph{arXiv preprint arXiv:2011.06801}.

\bibitem[{Juslin and Sloboda(2013)}]{juslin2013music}
Juslin, P.~N.; and Sloboda, J.~A. 2013.
\newblock \emph{Music and emotion.}
\newblock Elsevier Academic Press.

\bibitem[{Leung(2011)}]{leung2011comparison}
Leung, S.-O. 2011.
\newblock A comparison of psychometric properties and normality in 4-, 5-, 6-, and 11-point Likert scales.
\newblock \emph{Journal of social service research}, 37(4): 412--421.

\bibitem[{Ren et~al.(2020)Ren, He, Tan, Qin, Zhao, and Liu}]{ren2020popmag}
Ren, Y.; He, J.; Tan, X.; Qin, T.; Zhao, Z.; and Liu, T.-Y. 2020.
\newblock Popmag: Pop music accompaniment generation.
\newblock In \emph{Proceedings of the 28th ACM international conference on multimedia}, 1198--1206.

\bibitem[{Russell(1980)}]{russell1980circumplex}
Russell, J.~A. 1980.
\newblock A circumplex model of affect.
\newblock \emph{Journal of personality and social psychology}, 39(6): 1161.

\bibitem[{Stewart et~al.(2019)Stewart, Garrido, Hense, and McFerran}]{stewart2019music}
Stewart, J.; Garrido, S.; Hense, C.; and McFerran, K. 2019.
\newblock Music use for mood regulation: Self-awareness and conscious listening choices in young people with tendencies to depression.
\newblock \emph{Frontiers in psychology}, 10: 1199.

\bibitem[{Sulun, Davies, and Viana(2022)}]{sulun2022symbolic}
Sulun, S.; Davies, M.~E.; and Viana, P. 2022.
\newblock Symbolic music generation conditioned on continuous-valued emotions.
\newblock \emph{IEEE Access}, 10: 44617--44626.

\bibitem[{Tan and Herremans(2020)}]{tan2020music}
Tan, H.~H.; and Herremans, D. 2020.
\newblock Music fadernets: Controllable music generation based on high-level features via low-level feature modelling.
\newblock \emph{arXiv preprint arXiv:2007.15474}.

\bibitem[{Wallis et~al.(2011)Wallis, Ingalls, Campana, and Goodman}]{wallis2011rule}
Wallis, I.; Ingalls, T.; Campana, E.; and Goodman, J. 2011.
\newblock A rule-based generative music system controlled by desired valence and arousal.
\newblock In \emph{Proceedings of 8th international sound and music computing conference (SMC)}, 156--157.

\bibitem[{Wang et~al.(2024)Wang, Zhao, Liu, Pang, Qin, and Wu}]{wang2024review}
Wang, L.; Zhao, Z.; Liu, H.; Pang, J.; Qin, Y.; and Wu, Q. 2024.
\newblock A review of intelligent music generation systems.
\newblock \emph{Neural Computing and Applications}, 36(12): 6381--6401.

\bibitem[{Zhu et~al.(2018)Zhu, Liu, Yuan, Qin, Li, Zhang, Zhou, Wei, Xu, and Chen}]{zhu2018xiaoice}
Zhu, H.; Liu, Q.; Yuan, N.~J.; Qin, C.; Li, J.; Zhang, K.; Zhou, G.; Wei, F.; Xu, Y.; and Chen, E. 2018.
\newblock Xiaoice band: A melody and arrangement generation framework for pop music.
\newblock In \emph{Proceedings of the 24th ACM SIGKDD international conference on knowledge discovery \& data mining}, 2837--2846.

\end{thebibliography}

\end{document}